\documentclass[runningheads]{llncs}
\usepackage[T1]{fontenc}
\usepackage{graphicx}
\usepackage{algorithm}
\usepackage{algorithmic}
\usepackage{amsmath}
\usepackage{amsfonts}
\usepackage{booktabs}
\usepackage{tabularx}
\usepackage[hidelinks]{hyperref}
\usepackage{multirow}
\usepackage[table]{xcolor}
\usepackage{colortbl}
\usepackage{lipsum}
\usepackage{subcaption}

\newcommand{\prg}[1]{
\noindent
\textbf{#1}.}

\begin{document}
\title{CSC: Turning the Adversary’s Poison against Itself}
%
%
\author{Yuchen Shi$^\spadesuit$\inst{1}\orcidID{0009-0003-9656-4336} \and
Xin Guo\inst{1}\orcidID{0009-0002-8946-4188} \and
Huajie Chen$^\diamondsuit$\inst{1}\orcidID{0009-0009-8489-9386} \and
Tianqing Zhu \inst{1}\orcidID{0000-0003-3411-7947} \and 
Bo Liu \inst{2}\orcidID{0000-0002-3603-6617} \and 
Wanlei Zhou \inst{1}\orcidID{0000-0002-1680-2521}
}
\authorrunning{Yuchen et al.}
%
\institute{
Faculty of Data Science, City University of Macau, Taipa, Macau.
\email{\{D23090120503,H23090104001,hjchen,tqzhu,wlzhou\}@cityu.edu.mo}\\
 \and
Faculty of Engineering and Information Technology, University of Technology Sydney, New South Wales, Australia.
\email{Bo.Liu@uts.edu.au}\\
$^\spadesuit$First Author, $^\diamondsuit$Corresponding Author
}
\maketitle              
\begin{abstract}
Poisoning-based backdoor attacks pose significant threats to deep neural networks by embedding triggers in training data, causing models to misclassify triggered inputs as adversary-specified labels while maintaining performance on clean data.
Existing poison restraint-based defenses often suffer from inadequate detection against specific attack variants and compromise model utility through unlearning methods that lead to accuracy degradation.
This paper conducts a comprehensive analysis of backdoor attack dynamics during model training, revealing that poisoned samples form isolated clusters in latent space early on, with triggers acting as dominant features distinct from benign ones. 
Leveraging these insights, we propose Cluster Segregation Concealment (CSC), a novel poison suppression defense.
CSC first trains a deep neural network via standard supervised learning while segregating poisoned samples through feature extraction from early epochs, DBSCAN clustering, and identification of anomalous clusters based on class diversity and density metrics.
In the concealment stage, identified poisoned samples are relabeled to a virtual class, and the model's classifier is fine-tuned using cross-entropy loss to replace the backdoor association with a benign virtual linkage, preserving overall accuracy.
CSC was evaluated on four benchmark datasets against twelve poisoning-based attacks, CSC outperforms nine state-of-the-art defenses by reducing average attack success rates to near zero with minimal clean accuracy loss.
Contributions include robust backdoor patterns identification, an effective concealment mechanism, and superior empirical validation, advancing trustworthy artificial intelligence.

\end{abstract}

\section{Introduction}

Deep learning has been extensively utilized in diverse applications, including safety-critical domains where reliability is crucial, such as autonomous vehicles \cite{kuutti2020survey}, medical diagnosis \cite{shen2017deep}.
The training of deep learning models typically requires a substantial volume of training data.
However, acquiring large-scale, high-quality datasets presents significant challenges in practice.
Consequently, certain individuals or organizations may opt to source data from unreliable third-party platforms.

Although untrustworthy data sources offer a convenient means for developers to construct their deep learning models, they also introduce substantial security risks.
One prevalent threat is the poisoning-based backdoor attack, in which adversaries embed triggers into a subset of benign samples and reassign them to a target label specified by the adversary.
By subsequently training a deep learning model on this compromised dataset, the association 
between the backdoor trigger and the target label, namely the backdoor, is embedded within the model. 
During the inference stage, the backdoored model performs correctly on pure inputs but classifies any samples containing the backdoor trigger as the adversary-specified label.

Researchers have devised a range of empirical defenses to address the vulnerabilities posed by backdoor attacks.
Poison restraint-based approaches stand out among them, having shown encouraging outcomes in recent studies. 
Such defenses seek to thwart the establishment of backdoors through the mitigation of poisoned samples' influence amid model training.
As illustrations, Anti-Backdoor Learning (ABL) \cite{li2021antibackdoor} identifies poisoned from clean samples via loss examination and removes backdoors by applying machine unlearning to those poisoned samples; Decoupling-based Backdoor Defense (DBD) \cite{huang2022backdoor} thwarts backdoor threats by leveraging separate optimization of the feature extractor and classifier components; and NON-LinEarity (NONE) \cite{wang2022training} imposes non-linear boundaries in decision spaces during optimization to diminish backdoor efficacy.

Despite the proven efficacy of current poison restraint-based defenses, they encounter several shortcomings. 
In particular, the detection criteria employed by these methods can prove inadequate against specific attack variants \cite{chen2026turning}.
For example, DBD \cite{huang2022backdoor} relied on symmetric cross entropy loss which was initially intended for noisy label learning to identify poisoned samples, which may lead to ineffectiveness in countering clean-label attacks \cite{zhu2019transferable} since poisoned and benign samples display comparable symmetric cross entropy losses in these scenarios.
Such detection shortcomings subsequently compromise the backdoor elimination stage; misclassifying benign samples as poisoned, for instance, not only leaves backdoors intact but also degrades overall model utility.
Moreover, existing poison restraint-based defenses predominantly eradicate backdoors via machine unlearning \cite{wang2019neural,bourtoule2021machine,10.1145/3603620,chen2025queen,shi2026osmosis,chen2026hide}, which involves amplifying discrepancies between model outputs and labels of poisoned data.
While effective at backdoor mitigation these techniques frequently yield diminished classification precision or, in extreme cases, produce entirely dysfunctional models.
To overcome these challenges, we introduce \textit{CSC} a new defense approach that demonstrates roubust performance against poisoning-based backdoor attacks while preserving elevated accuracy in the resulting models.

\prg{Investigations} We begin comprehensively examining the fundamental processes underlying poisoning-based backdoor attacks.
This entails a detailed investigation into the dynamics of these attacks throughout the training of models, particularly in the latent representation space.
Our findings uncover important revelations.
Initially, the backdoor trigger serves as a dominant characteristic associated with the intended label, thereby facilitating quicker acquisition of poisoned samples relative to benign ones.
Consequently, this results in poisoned samples aggregating into a conspicuous segregated group, readily separable from benign samples during the early phases of model training.
Additionally, poisoned samples exhibit distinctive attributes in contrast to their benign counterparts.
By extension, a singular linkage emerges between backdoor attributes and the designated labels, markedly diverging from the connections linking benign attributes to the corresponding labels of benign samples.

\prg{Our Work} Drawing from the aforementioned investigations, we introduced an approach termed "Cluster Segregation Concealment (CSC)", which leverages the fundamental dynamics of data-poisoning backdoor attacks to counteract their impact and effectively ensnare the adversaries.
In detail, CSC initiates the process with conventional supervised training of a Deep Neural Network (DNN) model.
Concurrently, the module for isolating poisoned samples examines select early epochs to pinpoint contaminated data in the training set, achieving this by grouping example features across those epochs and pinpointing anomalous clusters likely harboring poisoned instances.
Following the conclusion of initial training, the backdoor concealment (BC) module refines the model's classifier component to excise embedded backdoors.
This involves initially assigning a novel virtual category to the identified poisoned samples, thereby forming an updated training set, followed by optimization of the classifier on this revised dataset via cross-entropy loss.
Consequently, model backdoors are neutralized through the establishment of an alternative virtual linkage that supplants the prior association between backdoor attributes and the intended labels.

We evaluate the defensive efficacy of CSC via comparisons against nine state-of-the-art (SOTA) poison restraint-based methods, utilizing four datasets under twelve distinct poison-based backdoor attack methods.
Our experimental findings reveal that CSC provides concurrent robust and efficient safeguarding.
Notably, it reduces the mean attack success rate to an extremely low level while imposing a negligible accuracy penalty on average, substantially surpassing the performance of alternative defenses.
Our main contributions are as follows:
\begin{itemize}
    \item We undertake a comprehensive examination of the dynamics in poisoning-based backdoor attacks throughout the model training stage, identifying robust patterns across diverse attack methods that highlight pronounced disparities in latent representations and label association trajectories between poisoned and benign samples.
    \item We introduce CSC, an innovative defense mechanism grounded in poison suppression, which proficiently eradicates backdoors from compromised models while upholding their classification efficacy.
    \item We validate CSC's effectiveness via rigorous testing against twelve distinct attack scenarios on four benchmark datasets, with outcomes indicating superior defensive outcomes relative to nine SOTA poison restraint-based alternatives.
\end{itemize}

\section{Related Work and Preliminaries}

\begin{table}[!pt]
    \scriptsize
    \centering
    \caption{Notations}
    \label{tab: notations}
    \begin{tabular}{c|l}
        \toprule Symbols      & Definitions                             \\
        \midrule 
        $\mathcal{T}$         & Trigger insertion mapping               \\
        $\mathcal{D}$         & Training dataset                        \\
        $\mathcal{D}_b$       & Benign training dataset                 \\
        $\mathcal{D}_p$       & Poisoned training dataset               \\
        $\mathcal{D}_{sp}$    & Segregated poisoned dataset             \\
        $\mathcal{D}_{sb}$    & Segregated benign dataset               \\
        $\mathcal{D}_{sp}^\dagger$ & Relabeled poisoned dataset         \\
        $\mathcal{D}^\dagger$  & Augmented dataset             \\
        $\mathcal{M}$         & Model                                   \\
        $h$                   & Classification head                     \\
        $\epsilon$            & Feature extractor                       \\
        $z$                   & Features                                \\
        \bottomrule
    \end{tabular}
\end{table}

\subsection{Deep Learning-Based Classifier}
Our study focuses on Deep Neural Network (DNN) models designed for image classification.
Consider an input space $\mathbb{X} \in \mathbb{R}^{H\times W \times C}$ with W denoting image width, H representing height and C indicating the number of channels, alongside a label set $\mathbb{Y} = \{1 ,2,3,  \dots,N\}$, where N specifies the total classes indexed by n.
A DNN for this purpose constitutes a prediction function $f: \mathbb{X} \rightarrow \mathbb{Y}$, whereby $f$ processes an input image $x \in \mathbb{X}$ to yield its corresponding class $y \in \mathbb{Y}$.
We decompose $f$ as $f = h \circ \epsilon $, with $h$ serving as the linear layer that converts latent embeddings to output predictions and $\epsilon$ as the feature extractor.
Acquiring such a $f$ typically demands assembling a training set $\mathcal{D}= \{(x_i,y_i)\}^n_{i=1}$, comprising image-label pairs.
Further, $f$ is optimized on $\mathcal{D}$ through the minimization of training loss $\mathbf{E}(f)$ governed by a specified loss $\mathcal{L}$:
\begin{equation}
    \mathbf{E}(f) = \frac{1}{n} \sum^n_{i=1} \mathcal{L}(f(x_i),y_i),
\end{equation}
where loss function ($\mathcal{L}$) is usually the vanilla cross-entropy (CE) loss in classification tasks.

\subsection{Poisoning-based Backdoor Attacks}

\subsubsection{Dirty-label attacks} represent a category of backdoor strategies where the designated target label diverges from the true label of the poisoned sample.
In these attacks, the adversary modifies a subset of the training data by injecting a trigger pattern and altering the labels to a specific target class, formally expressed as: given a dataset $\mathcal{D}=\{ (x_i,y_i)_{i=1}^n \}$, the poisoned dataset $\mathcal{D}_p$ includes samples $(x'_i = x_i + \delta, y'_i=t)$ where $t \neq y_i$ and $\delta$ denotes the trigger perturbation.
For example, Gu et al. proposed BadNets \cite{gu2017badnets}, a method to backdoor machine learning models using a blank pixel as a trigger to misclassify backdoor inputs as target labels.
Building upon this foundation, Liu et al. proposed the Trojan attack \cite{liu2018trojaning}, which inverts the neural architecture to craft a versatile trojan trigger, subsequently incorporating it into images to execute the assault. Similarly, Blend \cite{chen2017targeted} integrates triggers by merging arbitrary noise or Hello-Kitty visuals with the source image, enhancing concealment through blend ratio modulation.
In turn, Adap-Blend \cite{qi2023revisiting} adopts an adaptable methodology that avoids uniform label modifications across poisoned samples, utilizing a subdued trigger in the training stage contrasted with an intensified version at inference to ensure heightened evasion.

\subsubsection{Clean-label attacks} constitute a form of backdoor strategy wherein the assigned target label aligns with the true label of the poisoned samples.
In these paradigm, the adversary modifies a subset of the training data belonging to the target class $t$ by injecting a trigger perturbation while preserving the original labels, the poisoned dataset includes samples $(x'_i = x_i + \delta, y_i = t)$, where $x_i$ belongs to class $t$ and $\delta$ denotes the trigger perturbation.
For example, Barni et al. \cite{barni2019new} incorporate triggers via a ramp imposition on clean target-class samples, while Turner et al. \cite{turner2019label} alter select clean samples from the class using adversarial perturbations or generative architectures.

\subsubsection{Feature space attacks} represent a class of poisoning-based backdoor attacks in which the trigger is input-dependent, ensuring that distinct poisoned samples incorporate unique triggers.
Formally, such attacks employ a parameterized generator $( g(\cdot; \theta) : \mathcal{X} \to \mathcal{P}$ that produces a trigger $t = g(x; \theta)$ conditioned on the clean input $x \in \mathcal{X}$.
The poisoned sample is then constructed as $\hat{x} = B(x, t)$, where $B$ is an injection function (e.g., blending or warping), and the model is trained to map $\hat{x}$ to a target label while preserving normal behavior on clean inputs.

Representative examples include ABS \cite{10.1145/3319535.3363216}, which embeds triggers via input-specific image style modifications; IAB \cite{nguyen2020input}, which optimizes an input-aware generator to produce diverse, non-reusable triggers; WaNet \cite{nguyen2021wanet}, which exploits image warping for imperceptible, input-varying perturbations; and SSBA \cite{li2022baat,9711191}, which leverages a pretrained encoder-decoder to generate sample-specific triggers inspired by DNN-based steganography \cite{10.1145/3694965,chen2024high}.
These methods enhance stealthiness by breaking the uniform-trigger assumption underlying many defenses.

\subsection{Defense against Backdoor Attacks}
Numerous defenses counteract poisoning-based backdoor attacks in deep neural networks (DNNs), primarily using poison restraint methods that mitigate the influence of contaminated samples during training. ABL exploits the faster loss convergence of poisoned data. It identifies contaminants via loss trajectories and eradicates backdoors through unlearning, maximizing prediction errors on detected poisons against target classes. Similarly, DBD uses self-supervised pretraining to obtain a feature extractor insulated from trigger-target correlations. It optimizes the classifier using symmetric cross-entropy (SCE) loss, isolates benign samples with lower SCE values, and refines the model semi-supervisedly by assigning labels to clean data and treating the remainder as unlabeled.

ASD extends this approach by dynamically partitioning the dataset with SCE. It delineates hard clean samples from poisons via loss thresholds and applies semi-supervised learning across labeled and unlabeled pools to improve resilience. EBD exploits representational inconsistencies in poisoned images under transformations to stratify data into poisoned, benign, and ambiguous subsets. It introduces two schemes. D-BR performs iterative unlearning and relearning similar to ABL, whereas D-ST uses mixed cross-entropy training to learn on benign data while unlearning poisons.

NONE identifies backdoor-induced hyperplanes in neuron activations, preemptively filtering poisons to avoid these linear decision surfaces and yield backdoor-free models. Finally, CBD forgoes explicit detection. It harnesses a deliberately backdoored auxiliary model to capture causal attack effects and enforce representational independence during training, redirecting the focus to authentic causal pathways and inhibiting backdoor embedding.

\section{Problem Formulation}
\subsection{Constraints of current defense mechanisms}
Although existing poison suppression defenses against backdoor attacks in deep neural networks are effective, they exhibit limitations in detection robustness and model performance. Specific attack variants can evade current detection mechanisms. For example, ABL relies on loss reduction rates, failing in multi-target scenarios like BadNets. The SCE loss used by DBD and ASD targets noisy labels and proves inadequate against clean-label attacks. EBD assesses feature consistency under transformations, but this is circumvented when poisoned features remain invariant, as seen in CL. Additionally, NONE assumes a single trigger, failing against sample-specific attacks like SSBA.
Furthermore, these defenses often compromise model utility. ABL and EBD induce performance degradation via machine unlearning, risking loss explosion and catastrophic forgetting. DBD and ASD discard valuable labels in self-supervised or semi-supervised schemes. NONE impairs convergence through neuron resetting. Finally, CBD enforces representational independence, which may inadvertently limit the clean model's capacity by overlooking shared legitimate causal relationships.

\begin{figure}[t!]
\centering
\includegraphics[width=0.54\linewidth]{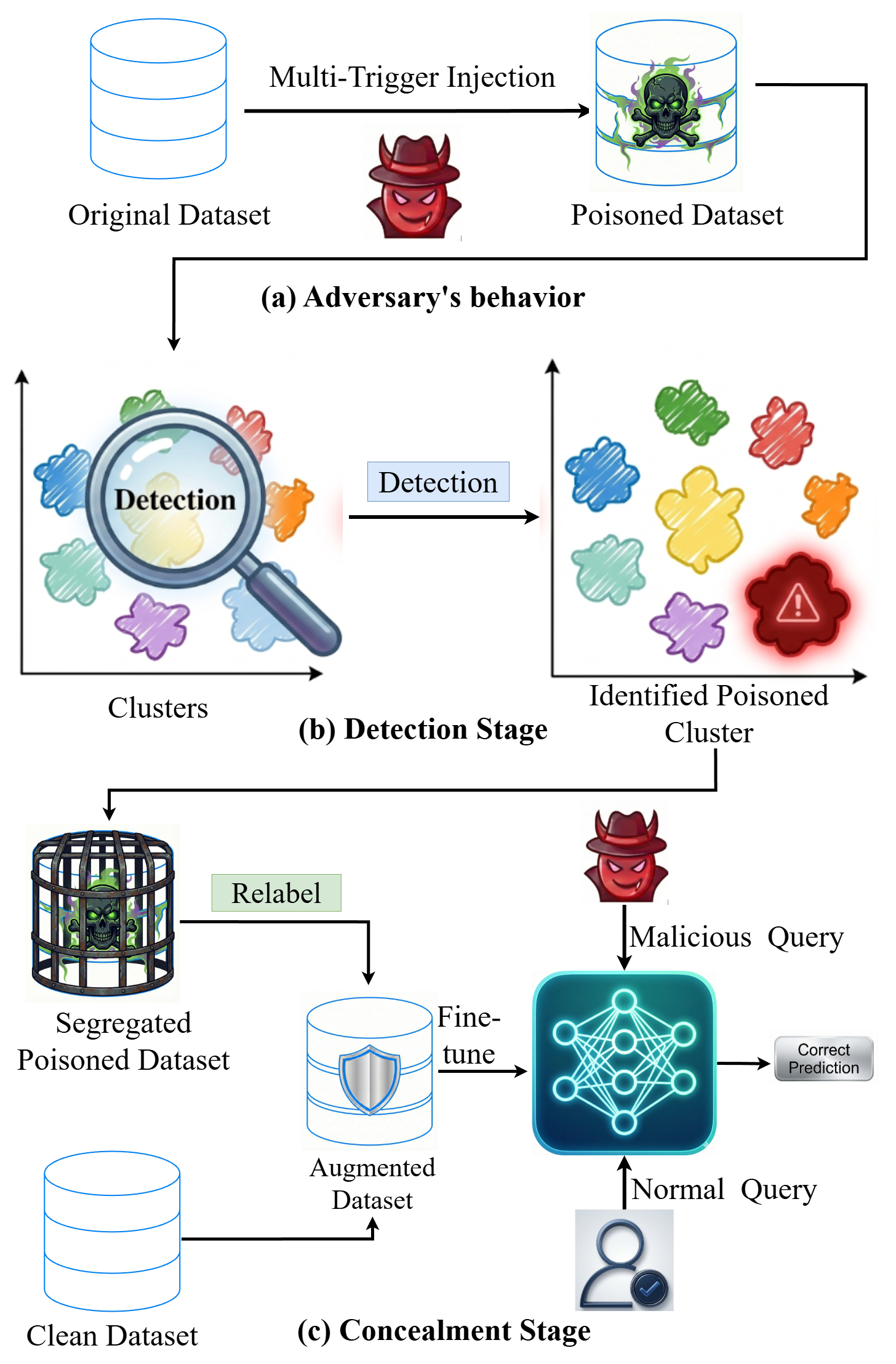}
\caption{Overview of CSC. (a) depicts the adversary's procedure for embedding a trigger, laying the groundwork for detecting poisoned samples within poisoning-based backdoor scenarios. (b) outlines the initial training stage, in which the model extracts features, performs clustering, and detects groupings indicative of potentially poisoned samples. (c) describes the backdoor concealment stage, during which we fine-tune the classifier. Initially we freeze the feature extractor and relabel the poisoned samples, followed by retraining the classification head to achieve the elimination of the backdoor.}
\label{fig: overview}
\end{figure}

\subsection{Threat Model}
\prg{Adversary's Goals} Adversaries seek to embed poisoned samples within a training set, enabling the result model to fulfill two goals: \textbf{G1: Efficacy.} The efficacy goal mandates that the model classify all inputs bearing a trigger selected by the adversary as the designated target class. \textbf{G2: Preservation.} The preservation goal ensures sustained elevated performance on untriggered input.

\prg{Adversary's Knowledge} In this study, we posit that adversaries have no access to or control over the training process, including the loss functions and model architectures.

\prg{Adversary's Capabilities} Adversaries are restricted to modifying a small proportion of the training data.
Formally, denote the benign training set as $\mathcal{D}_b = \{ (x_i, y_i) \}^n_{i=1}$, the adversary-defined trigger insertion mapping as $\mathcal{T}: \mathbb{X} \Rightarrow \tilde{\mathbb{X}}$ and the label reassigning mapping as $\zeta: \mathbb{X} \odot \mathbb{Y} \Rightarrow \tilde{\mathbb{Y}} $. 
Adversaries craft $\mathcal{T}$ and $\zeta$ to corrupt the subset of samples under their influence.
In detail, they covert a subset $\mathcal{D}_\upsilon \in \mathcal{D}_b$ into $\mathcal{D}'_{\upsilon} = \{ (\mathcal{T}(x_i), \zeta(x_i, y_i)) | (x_i, y_i \in \mathcal{D}_\upsilon)\}$, forming the poisoned training set $\mathcal{D}_p = \mathcal{D}'_\upsilon \cup (\mathcal{D}_b \backslash \mathcal{D}_\upsilon)$.
Upon training a model $\mathcal{M}$ using the poisoned dataset $\mathcal{D}_p$, the backdoor becomes embedded, leading model $\mathcal{M}$ to designate the adversary-chosen target label for any triggered inputs while maintaining elevated accuracy on untainted data.

\prg{Defender's Goal}
The goal of defender is to train a robust model using the provided dataset, such that inputs embedding a backdoor trigger are not classified as the adversary-specified label, while simultaneously preserving elevated accuracy on benign samples lacking any trigger.

\prg{Defender's Knowledge}
We consider a setting in which defenders maintain full authority over the training pipeline, including the selection of model architectures and loss functions. 
In contrast, defenders possess no prior knowledge regarding the proportion or distribution of poisoned samples within the provided training dataset.

\subsection{Backdoor Properties}
Backdoor attacks aim to implant triggers in a model so that any input containing the triggers is misclassified to a predetermined target label.
In essence, the attacker seeks to embed a spurious association between the trigger pattern and the target label within the model.
Poisoning-based backdoor attacks achieve this by deliberately inserting the trigger into a subset of training images and reassigning their labels to the target class.
The resulting poisoned samples exhibit two key properties:
\begin{itemize}
    \item \textbf{P1:} The trigger functions as a dominant feature predictive of the target label.
    High attack success rates require the model to establish a reliable mapping from the trigger to the target label.
    Consequently, the trigger must appear as a salient attribute naturally associated with the target class, thereby allowing the model to readily learn this robust association.
    \item \textbf{P2:} Poisoned samples possess distinctive features relative to clean samples.
    The strong induced correlation between the trigger and the target label typically produces highly discriminatory features in the poisoned data.
    Furthermore, for triggered inputs to be consistently classified as the target, the backdoor features must override the original semantic features of the samples.
    Thus, poisoned samples display feature representations that markedly differ from those of benign samples.
\end{itemize}

\section{Methodology}

\begin{figure}[!t]
    \centering
    \includegraphics[width=0.5\linewidth]{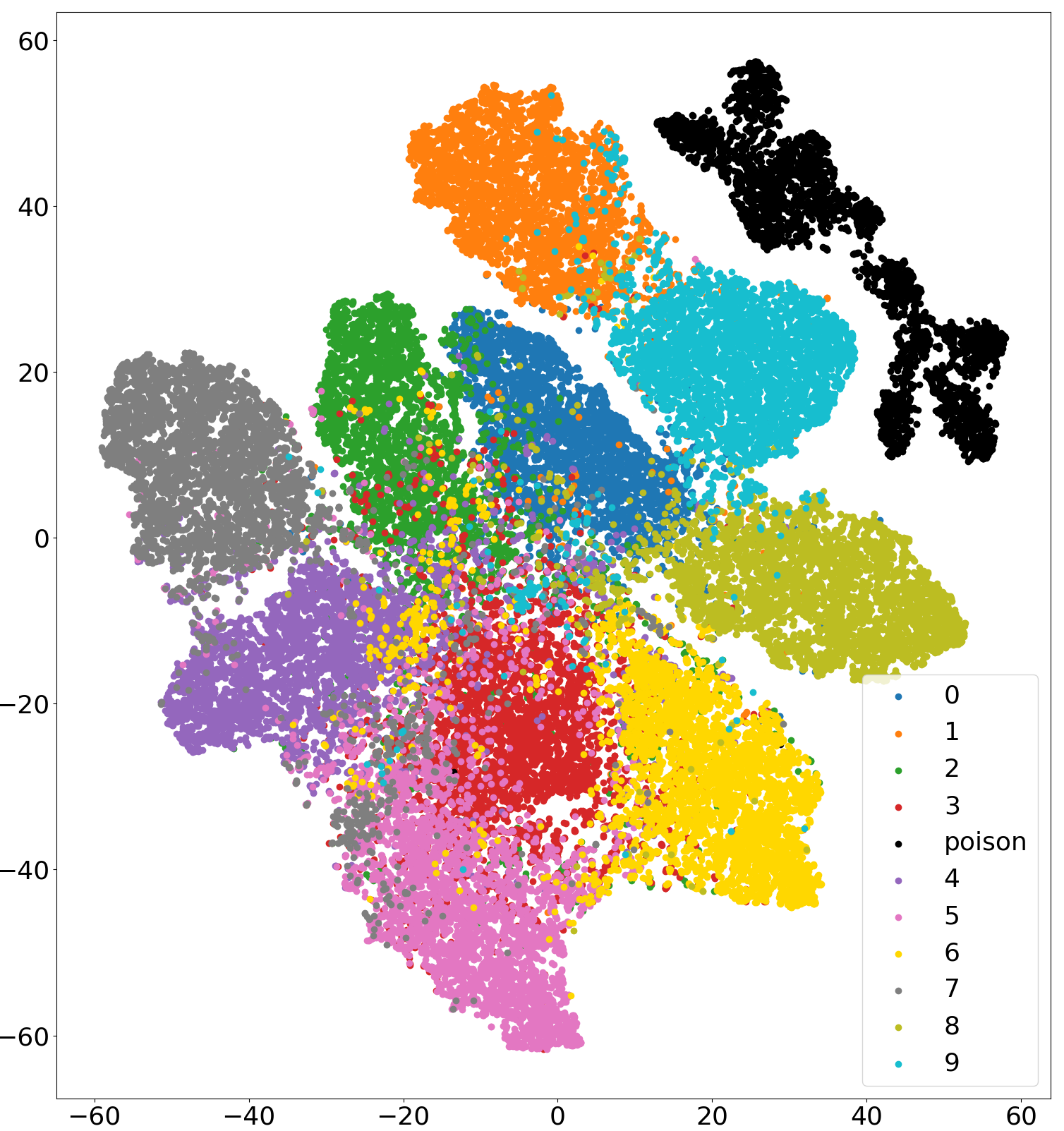}
    \caption{Visualization of training samples in latent space at the early training stage, the poisoned samples are shown in black cluster. }
    \label{fig: visualization}
\end{figure}

The CSC encompasses two primary stages (An overview of CSC is presented in \autoref{fig: overview}):  (1) Segregation and (2) Concealment.
During the Segregation stage, the model $\mathcal{M}$ undergoes training on the training dataset $\mathcal{D}$ through standard supervised learning.
Within this training procedure, poisoned samples isolation is applied by examining early training epochs to identify poisoned samples in $\mathcal{D}$.
The concealment stage involves executing backdoor hiding to eliminate the vulnerabilities embedded in the model from prior stage.
Specifically, the segregated poisoned subset $\mathcal{D}_{sp}$ is reassigned labels, after which the classification component of the model is retrained on $\mathcal{D}_{sp}$ to obscure the backdoors.
Additional details on CSC are provided in the subsequent sections.

\subsection{Segregation Stage}
Segregation stage seeks to identify poisoned samples within the provided training dataset $\mathcal{D}$.
We posit that, in the preliminary stage of model training, these poisoned samples will coalesce into a notable anomalous cluster, readily separable from benign counterparts.
Thus, our approach involves scrutinizing the representations of training samples across the $EP_{detct}$ epochs to differentiate poisoned from benign samples.
As illustrated in \autoref{fig: overview}, the process for this segregation method initiates by deriving latent features from samples $x$ in dataset $\mathcal{D}$.
These representations are subsequently partitioned into multiple clusters.
Thereafter, we assess the clusters to locate the aberrant group likely encompassing poisoned samples.
Poisoned samples are ultimately pinpointed using these anomalous clusters.

\subsubsection{Feature Extraction:} This process commences with the derivation of representations for samples within the given training set $\mathcal{D}$.
Consistent with prior studies \cite{wang2022training,chen2022effective}, our approach utilizes representations of the second to last layer, extracted for the training samples.
Specifically, for each sample $x_i \in \mathcal{D}$, we forward it through the model $\mathcal{M}$ and capture the second to last layer's output as its corresponding feature $z_i$.
To formalize this, we decompose $\mathcal{M} = h \circ \epsilon $, with $\epsilon$ denoting the feature extractor comprising all layers preceding the final one in $\mathcal{M}$, and $h$ representing the classification head, embodied by $\mathcal{M}$'s terminal fully-connected layer.
Accordingly, the representation $z_i$ for $x_i \in \mathcal{D}$ is expressed as $z_i = \epsilon(x_i)$.

\subsubsection{Clustering:}
We utilize the density-based spatial clustering of applications with noise (DBSCAN) \cite{10.5555/3001460.3001507} to partition the latent representations $z$ derived from all training samples into separate clusters.
Our rationale for selecting DBSCAN encompasses two key aspects: (I) Establishing the cluster count within the representation space presents a considerable challenge, thereby disqualifying approaches like K-means that demand advance specification of this quantity. 
Instead, as a density-oriented method, DBSCAN aggregates spatially adjacent points in regions of elevated density, obviating the need for predefined cluster numbers.
(II) Furthermore, DBSCAN exhibits established efficacy and efficiency, could with extensive application in diverse fields.
Clustering via DBSCAN involves two hyperparameters: \textbf{eps}, which specifies the threshold distance delineating neighborhood extents for data points, and \textbf{MinPts}, which indicates the requisite number of adjacent points to constitute a cluster.

\subsubsection{Suspicious Cluster Identification:}
\label{section: Suspicious Cluster Identification}
Utilizing the obtained feature groupings, we proceed to detect anomalous clusters potentially incorporating poisoned samples.
The underlying principle deems a cluster suspicious provided it does not constitute the most populous group.
In line with this, we sequentially evaluate each group generated via the DBSCAN method, assessing its suspicious status through metrics capturing the diversity of classes it spans alongside its density.
A cluster qualifies as suspicious if it spans either a single class or the full set of classes while simultaneously not tanking as the largest.

\begin{algorithm}[!t]
\caption{Poisoned Sample Segregation}
\label{alg: poisoned samples segregation}
\begin{algorithmic}[1]
\STATE \textbf{Input:} Dataset $D$, Classifier $\mathcal{M}$, Detection Epochs $EP_{detect}$ 
\STATE \textbf{Output:} Segregated Poisoned Dataset $\mathcal{D}_{sp}$, Segregated Benign Dataset $ \mathcal{D}_{sb}$, Model $\mathcal{M}$ 
\STATE allSuspiciousClusters $\leftarrow \emptyset$;
\WHILE{Epoch $e \leq  EP_{detct}$}
\STATE Training the classifier $\mathcal{M}$;
\STATE $z_i$ $\leftarrow$ ExtractFeature $(\mathcal{M}, D)$;
\STATE $z_i$ $\leftarrow$ t-SNE ($z_i$);
\STATE clusters $\leftarrow$ DBScan ($z_i$, $D$);
\STATE sc $\leftarrow$ DetectClusters (clusters);
\STATE asc $\leftarrow$ allSuspiciousClusters.Extend (sc);
\ENDWHILE
\STATE $ \mathcal{D}_{sp} \leftarrow$ asc;
\STATE $ \mathcal{D}_{sb} \leftarrow D \setminus {D}_{sp}$;
\RETURN $\mathcal{D}_{sp}, \mathcal{D}_{sb}, \mathcal{M}$
\end{algorithmic}
\end{algorithm}
\subsubsection{Poisoned Sample Identification:} Leveraging the identified anomalous clusters, we ultimately pinpoint the poisoned samples embedded in the training set from among these groups.
The procedure for detecting such poisoned samples is delineated in Alg. \autoref{alg: poisoned samples segregation}.
Within this framework, $\mathcal{D}$ signifies the initial training dataset, potentially harboring poisoned samples; $\mathcal{M}$ represents the model slated for optimization and $EP_{detect}$ indicates the count of preliminary epochs utilized for isolating the poisoned samples.

We begin by conducting standard supervised learning of the model $\mathcal{M}$ for $EP_{detect}$ initial epochs, during which suspicious clusters are identified at the end of each epoch.
Specifically, within every epoch $e$, the model is first updated using conventional supervised optimization. Subsequently, feature representations are extracted from all training samples, followed by dimensionality reduction of these representations.
The reduced features are then grouped into distinct clusters.
Next, suspicious clusters are detected according to the criterion described in \autoref{section: Suspicious Cluster Identification}, and these clusters are accumulated in the global list "allSuspiciousClusters".
Per our established rule, any cluster other than the largest (i.e., the one encompassing the greatest number of samples) is classified as suspicious.
Ultimately, all samples residing in suspicious clusters are designated as poisoned samples, thereby constituting the poisoned subset $\mathcal{D}_{sp}$, while the original dataset is partitioned into benign and poisoned portions.

\prg{Note} In Alg. \autoref{alg: poisoned samples segregation}, poisoned samples within the training dataset are identified by aggregating information across a total of $EP_{detect}$ epochs rather than relying on a single epoch.
This design choice stems from the inherent instability of model training during its initial stages, which our poisoned sample isolation method specifically targets.
Consequently, detection based on any individual epoch tends to yield suboptimal performance owing to the stochastic nature of the optimization process.
By incorporating dynamics from all  $EP_{detct}$ epochs, we achieve more consistent training trajectories, thereby enhancing the effectiveness and reliability of poisoned samples detection.

\subsection{Concealment Stage}
Here, we present our Backdoor Concealment (BC) approach, designed to counteract the adverse effects induced by poisoned samples following the completion of standard supervised learning.
The core principle underlying this method involves disrupting the association between backdoor features and the adversary-selected target labels through retraining of the model's classification head $h$ within the overall architecture $\mathcal{M}$.

Utilizing the benign dataset $\mathcal{D}_{sb}$ and the poisoned dataset $\mathcal{D}_{sp}$ obtained through poisoned sample segregation, we initially reassign labels to the samples in $\mathcal{D}_{sp}$ by associating them with a novel confusion class.
This reassignment disrupts the linkage between the poisoned samples and the target label selected by the adversary.
In effect, we introduce an additional poisoning step to the poisoned dataset $\mathcal{D}_{sp}$ using this new confusion class.
Given that the original dataset $\mathcal{D}$ comprises $n$ classes, the confusion class is assigned the index $n + 1$. 
The resulting relabeled poisoned dataset is therefore defined as $\mathcal{D}_{sp}^\dagger = \{ (x, n+1) \mid x \in \mathcal{D}_{sp} \}$.
The augmented training set $\mathcal{D}^\dagger$ is subsequently formed by merging the benign dataset $\mathcal{D}_{sb}$ with this relabeled poisoned set $\mathcal{D}_{sp}^\dagger$, yielding $\mathcal{D}^\dagger = \mathcal{D}_{sb} \cup \mathcal{D}_{sp}^\dagger$.

Next, we embed the revised association linking the poisoned samples to the novel introduced confusion class $n + 1$ within the model.
This is accomplished by re-training solely the classification head $h$ of $\mathcal{M}$ for $E_{c}$ epochs on the augmented training dataset $\mathcal{D}^\dagger$, employing cross entropy loss.
Throughout this retraining stage, the feature extractor $\epsilon$ remains frozen, since it has already developed the capacity to differentiate backdoor features from benign ones, and our objective is confined to altering the correspondence between backdoor features and the adversary-selected target label.
Moreover, preserving the feature extractor $\epsilon$ in a fixed state enables the classification head $h$ to leverage discriminative representations more effectively previously acquired by $\epsilon$.
Consequently, this BC strategy successfully redirects backdoor features towards the confusion class $n + 1$, while preserving strong classification performance on benign samples.

\section{Experiments}

\begin{table}[!t]
\centering
\caption{Comparisons of CSC with nine SOTA defense strategies (\%). The results obtained with ResNet-18.}
\resizebox{\linewidth}{!}{
\begin{tabular}{l|l|cc|cc|cc|cc|cc|cc|cc|cc|cc|cc|cc}
\toprule
Dataset & Defense$\rightarrow$ & \multicolumn{2}{c|}{No defense} & \multicolumn{2}{c|}{DP-SGD \cite{Hong2020OnTE}} & \multicolumn{2}{c|}{NAD \cite{li2021neural}} & \multicolumn{2}{c|}{ABL \cite{li2021antibackdoor}} & \multicolumn{2}{c|}{NONE \cite{wang2022training}} & \multicolumn{2}{c|}{DBD \cite{huang2022backdoor}} & \multicolumn{2}{c|}{D-BR \cite{chen2022effective}} & \multicolumn{2}{c|}{D-ST \cite{chen2022effective}} & \multicolumn{2}{c|}{CBD \cite{10204451}} & \multicolumn{2}{c|}{ASD \cite{10204454}} & \multicolumn{2}{c}{\textbf{Ours}} \\ 

 & Attack$\downarrow$ & ACC & ASR & ACC & ASR & ACC & ASR & ACC & ASR & ACC & ASR & ACC & ASR & ACC & ASR & ACC & ASR & ACC & ASR & ACC & ASR & ACC & ASR \\ 
\hline
\multirow{12}{*}{CIFAR-10 \cite{Krizhevsky2009LearningML}} 
& No attack & 94.92 & 0.0 & 61.36 & 0 & 82.72 & 0.0 & 80.59 & 0.0 & 91.09 & 0.0 & 89.74 & 0.0 & 84.43 & 0 & 91.03 & 0.0 & 90.51 & 0.0 & 51.87 & 0 & 95.69 & 0.0 \\ 

 & BadNets \cite{gu2017badnets} & 93.13 & 89.19 & 50.89 & 3.57 & 78.96 & 3.12 & 78.92 & 75.45 & 92.78 & 21.92 & 90.83 & 3.38 & 81.59 & 79.41 & 86.54 & 10.98 & 91.57 & 85.69 & 90.34 & 12.03 & 94.98 & 0.0 \\ 
 
 & Trojan \cite{liu2018trojaning} & 94.73 & 100.0 & 52.68 & 95.42 & 82.98 & 3.57 & 87.01 & 0.23 & 93.29 & 2.12 & 92.84 & 2.38 & 90.92 & 0.89 & 94.12 & 0.0 & 90.78 & 1.98 & 92.98 & 1.69 & 94.63 & 0.0 \\ 
 
 & Blend-Strip \cite{chen2017targeted} & 94.86 & 100.0 & 51.46 & 100.0 & 82.91 & 54.75 & 82.45& 3.32 & 91.45 & 41.02 & 93.13 & 1.93 & 90.35 & 71.55 & 93.19 & 0.0 & 90.77 & 4.51 & 91.64 & 100.0 & 95.07 & 0.01 \\ 
 
 & Blend-Kitty \cite{chen2017targeted} & 94.62 & 100.0 & 53.18 & 99.40 & 84.11 & 0.75 & 83.25 & 0.0 & 92.54 & 2.83 & 92.24 & 2.33 & 92.04 & 64.56 & 93.23 & 0.15 & 90.87 & 47.78 & 92.57 & 0.68 & 94.23 & 0.0 \\ 
 
 & SIG \cite{barni2019new} & 94.79 & 99.35 & 52.78 & 78.15 & 87.31 & 48.45 & 69.17 & 0.37 & 93.65 & 99.24 & 90.55 & 23.89 & 79.24 & 0.0 & 89.9 & 95.26 & 90.39 & 6.35 & 92.18 & 98.31 & 93.67 & 0.03 \\ 
 
 & CL \cite{turner2019label} & 94.38 & 89.47 & 51.41 & 15.79 & 83.66 & 0.39 & 74.52 & 0.61 & 92.57 & 2.11 & 90.38 & 16.22 & 72.21 & 48.12 & 89.27 & 15.34 & 90.13 & 62.10 & 92.36 & 0.85 & 93.76 & 0.0 \\ 
 
 & WaNet \cite{nguyen2021wanet} & 95.34 & 97.79 & 53.67 & 13.22 & 86.56 & 4.30 & 76.43 & 5.01 & 91.25 & 20.31 & 91.43 & 2.98 & 88.43 & 80.22 & 88.43 & 20.32 & 91.06 & 12.59 & 90.28 & 1.68 & 94.77 & 0.13 \\ 
 
 & ABS \cite{10.1145/3319535.3363216} & 94.86 & 99.45 & 47.39 & 77.37 & 84.56 & 7.12 & 72.21 & 4.64 & 91.78 & 80.34 & 89.04 & 26.56 & 91.72 & 97.48 & 87.43 & 98.38 & 92.30 & 6.23 & 91.90 & 1.33 & 93.07 & 0.03 \\ 
 
 & IAB \cite{nguyen2020input} & 94.56 & 100.0 & 52.43 & 96.39 & 79.12 & 1.43 & 91.48 & 0.15 & 92.56 & 15.44 & 90.32 & 86.35 & 88.85 & 91.34 & 87.39 & 79.38 & 90.07 & 6.44 & 92.45 & 1.48 & 93.91 & 0.0 \\ 
 
 & Adap-Blend \cite{qi2023revisiting} & 95.02 & 90.29 & 52.38 & 15.59 & 84.43 & 20.73 & 61.39 & 90.39 & 92.45 & 67.04 & 91.24 & 75.65 & 84.24 & 63.28 & 90.37 & 66.66 & 91.70 & 77.0 & 92.41 & 28.04 & 94.34 & 0.0 \\ 
 
 & \textit{Avg.} & 94.66 & 96.55 & 52.69 & 59.49 & 83.39 & 14.46 & 77.94 & 18.01 & 92.31 & 35.24 & 91.07 & 24.17 & 85.82 & 59.69 & 90.08 & 38.65 & 90.92 & 31.07 & 88.27 & 24.61 & 94.37 & 0.02 \\ 
 
\hline
\multirow{8}{*}{CIFAR-100 \cite{Krizhevsky2009LearningML}}
& No attack & 75.51 & 0.0 & 43.59 & 0.0 & 60.85 & 0 & 68.19 & 0.0 & 69.72 & 0.0 & 69.53 & 0.0 & 70.96 & 0.0 & 73.87 & 0.0 & 71.79 & 0.0 & 67.59 & 0.0 & 74.78 & 0.0 \\ 

 & BadNets \cite{gu2017badnets} & 74.13 & 100.0 & 38.12 & 33.75 & 65.07 & 0.42 & 68.49 & 0.0 & 69.41 & 8.46 & 70.25 & 0.0 & 60.19 & 61.98 & 68.46 & 71.39 & 68.65 & 5.54 & 67.37 & 0.45 & 74.01 & 0.02 \\ 
 
 & Trojan \cite{liu2018trojaning} & 74.35 & 100.0 & 37.16 & 52.27 & 63.03 & 1.49 & 67.41 & 0.0 & 72.42 & 0.36 & 71.04 & 0.17 & 72.76 & 0.0 & 71.38 & 0.0 & 70.34 & 0.76 & 67.11 & 2.09 & 73.38 & 0.01 \\ 
 
 & Blend-Strip \cite{chen2017targeted} & 74.41 & 100.0 & 43.36 & 98.47 & 63.23 & 0.45 & 64.91 & 0.0 & 71.35 & 64.53 & 70.29 & 5.25 & 72.16 & 0.42 & 71.53 & 0.0 & 70.83 & 8.14 & 66.92 & 99.18 & 74.32 & 0.0 \\ 
 
 & Blend-Kitty \cite{chen2017targeted}& 74.18 & 100.0 & 43.71 & 96.15 & 65.62 & 6.38 & 68.18 & 0.0 & 73.12 & 40.27 & 70.25 & 22.67 & 71.13 & 0.0 & 71.39 & 0.0 & 70.18 & 21.86 & 68.16 & 99.34 & 74.15 & 0.0 \\ 
 
 & WaNet \cite{nguyen2021wanet} & 73.18 & 88.78 & 41.03 & 36.18 & 66.34 & 1.89 & 65.23 & 0.0 & 68.21 & 77.34 & 70.21 & 7.58 & 71.25 & 77.84 & 70.15 & 42.94 & 71.23 & 80.70 & 63.29 & 7.08 & 72.85 & 0.0 \\ 
 
 & ABS \cite{10.1145/3319535.3363216} & 75.30 & 100.0 & 40.17 & 88.05 & 62.62 & 0.71 & 65.62 & 0.0 & 70.96 & 75.53 & 72.56 & 8.33 & 72.13 & 99.76 & 71.05 & 97.37 & 70.22 & 55.12 & 68.12 & 64.36 & 73.29 & 0.01 \\ 
 
 & \textit{Avg.} & 74.44 & 98.13 & 41.02 & 67.48 & 63.82 & 1.89 & 66.86 & 0.0 & 70.74 & 44.41 & 70.59 & 7.40 & 70.08 & 40.0 & 71.12 & 35.28 & 70.46 & 28.69 & 66.94 & 45.42 & 73.83 & 0.01 \\
 
\hline
\multirow{10}{*}{GTSRB \cite{Stallkamp2012}}
& No attack & 98.34 & 0.0 &  59.21 & 0.0 &  95.31 & 0.0 & 95.75 & 0.0 & 95.61 & 0.0 & 92.34 & 0.0 & 96.73 & 0.0 & 95.91 & 0.0 & 95.26 & 0.0 & 94.28 & 0.0 & 97.24 & 0.0 \\ 

& BadNets \cite{gu2017badnets} & 97.43 & 100.0 & 55.23 & 89.32 & 91.64 & 0.27 & 95.25 & 0.0 & 92.41 & 1.21 & 92.24 & 0.0 & 95.76 & 0.35 & 92.32 & 2.91 & 92.56 & 0.57 & 91.63 & 0.0 & 97.14 & 0.0 \\ 

 & Trojan \cite{liu2018trojaning} & 97.57 & 98.48 & 53.13 & 90.46 & 93.45 & 1.34 & 93.32 & 0.28 & 95.58 & 0.69 & 93.18 & 0.77 & 96.43 & 4.38 & 95.16 & 0.0 & 94.23 & 0.21 & 94.36 & 0.10 & 97.23 & 0.0 \\ 
 
 & Blend-Strip \cite{chen2017targeted} & 97.22 & 100.0 & 56.29 & 21.23 & 90.58 & 17.44 & 93.75 & 10.39 & 95.35 & 2.87 & 90.18 & 3.35 & 95.32 & 30.54 & 92.81 & 0.19 & 95.74 & 0.83 & 90.44 & 0.10 & 96.37 & 0.21 \\ 
 
 & Blend-Kitty \cite{chen2017targeted} & 96.13 & 100.0 & 54.73 & 97.14 & 91.08 & 0.59 & 93.10 & 2.67 & 95.23 & 11.65 & 92.41 & 10.92 & 94.21 & 5.72 & 93.44 & 15.65 & 96.75 & 2.58 & 95.36 & 1.56 & 95.94 & 0.37 \\ 
 
 & SIG \cite{barni2019new} & 96.98 & 97.14 & 57.34 & 10.35 & 90.56 & 7.68 & 94.64 & 8.62 & 96.16 & 6.72 & 91.46 & 2.44 & 95.53 & 41.82 & 90.47 & 94.32 & 93.24 & 7.41 & 94.66 & 20.05 & 96.83 & 0.21 \\ 
 
 & CL \cite{turner2019label} & 97.26 & 99.34 & 57.66 & 1.97 & 93.53 & 24.14 & 94.02 & 9.18 & 95.35 & 0.14 & 94.26 & 6.37 & 93.29 & 5.38 & 95.34 & 4.24 & 96.83 & 4.93 & 95.29 & 1.43 & 96.36 & 0.14 \\ 
 
 & WaNet \cite{nguyen2021wanet} & 96.85 & 94.38 & 54.48 & 19.0 & 89.33 & 42.32 & 91.54 & 6.48 & 97.37 & 97.24 & 90.56 & 4.35 & 87.34 & 76.66 & 93.22 & 6.51 & 95.93 & 3.22 & 92.19 & 1.97 & 95.97 & 0.22 \\ 
 
 & IAB \cite{nguyen2020input} & 97.41 & 100.0 & 54.25 & 89.64 & 93.34 & 23.17 & 94.27 & 5.21 & 95.93 & 5.62 & 94.73 & 4.21 & 95.34 & 1.83 & 95.35 & 0.0 & 95.82 & 0.76 & 94.54 & 0.59 & 96.24 & 0.33 \\ 
 
 & \textit{Avg.} & 97.24 & 98.67 & 55.81 & 52.39 & 92.09 & 14.62 & 93.96 & 5.35 & 65.44 & 15.77 & 92.37 & 4.05 & 94.43 & 20.84 & 93.78 & 15.48 & 95.15 & 2.56 & 93.64 & 3.23 & 96.59 & 0.19 \\ 
\hline

\multirow{7}{*}{Tiny-ImageNet \cite{deng2009imagenet}}
& No attack & 87.61 & 0 & 45.38 & 0 & 87.01 & 0.0 & 83.86 & 0.0 & 85.28 & 0.0 & 76.41 & 0.0 & 84.22 & 0.0 & \multicolumn{2}{c|}{/} & 86.88 & 0.0 & 75.89 & 0.0 & 88.23 & 0.0 \\ 

 & BadNets \cite{gu2017badnets} & 87.19 & 95.23 & 43.19 & 30.25 & 88.11 & 52.35 & 80.12 & 0.10 & 86.19 & 0.78 & 77.91 & 2.95 & 86.16 & 0.19 & \multicolumn{2}{c|}{/} & 82.38 & 91.21 & 74.56 & 0.32 & 86.49 & 0.02 \\ 
 
 & Blend-Strip \cite{chen2017targeted} & 86.76 & 92.56 & 49.33 & 81.53 & 88.23 & 46.37 & 80.34 & 0.26 & 79.31 & 99.16 & 75.59 & 0.32 & 85.63 & 15.28 & \multicolumn{2}{c|}{/} & 84.54 & 94.61 & 75.36 & 0.42 & 87.12 & 0.0 \\ 
 
 & Blend-Kitty \cite{chen2017targeted} & 86.34 & 99.15 & 47.35 & 98.68 & 82.52 & 98.74 & 76.47 & 0.0 & 85.16 & 97.95 & 74.33 & 8.73 & 85.37 & 13.48 & \multicolumn{2}{c|}{/} & 85.34 & 65.30 & 75.33 & 5.22 & 85.92 & 0.0 \\ 
 
 & SSBA \cite{li2022baat} & 88.91 & 99.13 & 40.48 & 89.17 & 88.27 & 64.19 & 80.18 & 0.0 & 85.74 & 97.31 & 76.56 & 15.32 & 84.11 & 97.37 & \multicolumn{2}{c|}{/} & 86.10 & 74.64 & 75.16 & 1.34 & 87.93 & 0.0 \\ 
 
 & WaNet \cite{nguyen2021wanet}& 86.14 & 98.32 & 47.51 & 51.21 & 87.47 & 56.35 & 80.16 & 10.97 & 85.73 & 55.74 & 75.90 & 19.44 & 85.10 & 19.7 & \multicolumn{2}{c|}{/} & 83.22 & 16.12 & 73.13 & 0.69 & 85.33 & 0.28 \\ 
 
& \textit{Avg.} & 87.17 & 96.88 &  45.54 & 70.17 &  86.94 & 63.60 &  80.19 & 2.27 & 84.57 & 70.19 & 76.12 & 9.35 & 85.10 & 29.20  & \multicolumn{2}{c|}{/}  & 84.74 & 68.38 & 74.91 & 1.60 & 86.84 & 0.06 \\ 

\bottomrule
\end{tabular}
}
\label{tab: main results resnet18}
\end{table}

\subsection{Settings}
\prg{Datasets and Models} We comprehensively evaluate the proposed CSC method on four widely adopted benchmark datasets: CIFAR-10 \cite{Krizhevsky2009LearningML}, CIFAR-100 \cite{Krizhevsky2009LearningML}, GTSRB \cite{Stallkamp2012}, and Tiny-ImageNet \cite{deng2009imagenet}.
For model architectures, we employ ResNet-18 \cite{he2016deep} to assess its robustness.

\prg{Attacks} Our evaluation encompasses 12 poisoning-based backdoor attacks, comprising six dirty-label attacks (BadNets, Trojan, Blend-Strip, Blend-Kitty, and Adap-Blend), two clean-label attacks (CL and SIG), and four feature-level attacks (ABS, IAB, WaNet and SSBA).
In addition, we include the natural trigger setting in CIFAR-10.
All attacks are implemented using the publicly available source code provided with their respective publications, preserving their original hyperparameter configurations.
Moreover, certain attacks (such as clean-label attacks and Trojan) are excluded from CIFAR-100 and Tiny-ImageNet, as they do not scale effectively to these datasets.

\prg{Baselines} We evaluate our proposed CSC against 9 SOTA defense strategies: ABL, DBD, D-BR, D-ST, NONE, CBD, ASD, DP-SGD, and NAD.
For each of these 9 methods, we employ their publicly released official implementations along with the default hyperparameter settings reported in the respective original manuscripts.
In addition, we report results for standard deep neural networks trained in the absence of any defensive mechanism.

\prg{Evaluation Metrics} We adopt Accuracy (ACC) and Attack Success Rate (ASR) as the primary metrics for assessing defense effectiveness.
ACC measures the model's classification performance on the clean test set, whereas ASR quantifies the proportion of poisoned test samples correctly classified as the adversary's target label.
Higher ACC reflect superior classification capability of the DNN, while lower ASR indicate greater robustness against backdoor attacks.
In addition, we report precision and recall to evaluate the efficacy of poisoned sample detection.
Precision represents the fraction of isolated samples that are genuinely poisoned, and recall denotes the fraction of all poisoned samples successfully identified by the segregation procedure.

\prg{Hyperparameters and Implementation} The following hyperparameter settings are employed for our CSC method.
The standard supervised training stage spans $e=100$ epochs.
For the poisoned sample segregation component, we examine the feature representations across the first $EP_{detect} = 10$ epochs.
The DBSCAN clustering algorithm is configured with $eps = 3$.
In the BC stage, the classifier portion of the model is retrained for $E_c = 10$ epochs.
Our method was based on the PyTorch framework, and all experiments were conducted on single NVIDIA GeForce RTX 4090 GPU, running Ubuntu 24.04 LTS.

\subsection{Main Results}
Our comprehensive evaluation across multiple datasets demonstrates that the CSC framework provides robust protection against adversarial manipulations with minimal impact on model utility. Relative to undefended models, CSC substantially decreases the mean attack success rate (ASR) from baseline highs exceeding $96\%$ to near zero across CIFAR-10 ($0.02\%$), CIFAR-100 ($0.01\%$), GTSRB ($0.19\%$), and Tiny-ImageNet ($0.06\%$). This mitigation occurs alongside negligible average accuracy (ACC) declines of $0.29\%$, $1.18\%$, and $0.65\%$ on the first three datasets, respectively. Furthermore, CSC exhibits clear advantages over state-of-the-art baselines. While nine alternative defenses yield average ASRs ranging from $14.4\%$ to $54.49\%$ on CIFAR-10, CSC achieves a superior $0.02\%$. Similarly, on CIFAR-100 and GTSRB, defenses such as NONE, D-BR, D-ST, and DBD struggle to suppress ASRs effectively, leaving residual rates between $4.05\%$ and $44.41\%$. When baselines do achieve lower ASRs, they often incur considerable penalties on model utility. For example, NAD and ABL drop ACC by $11.9\%$ and $8.15\%$ on CIFAR-100, while CBD and ASD diminish ACC by $2.09\%$ and $3.6\%$ on GTSRB. In terms of ACC preservation, CSC consistently outperforms competitors. On CIFAR-10, it surpasses defenses attaining over 90\% ACC (NONE, DBD, D-ST, CBD, and ASD) by mean differences of $2.06\%$, $3.3\%$, $4.29\%$, $3.45\%$, and $3.46\%$, respectively. This scalability extends to the high-resolution Tiny-ImageNet dataset, where CSC yields average ACC improvements of $8.5\%$, $2.3\%$, $10.72\%$, and $11.93\%$ relative to ABL, NONE, DBD, and ASD. Finally, CSC reliably counters specific attack variants that exploit baseline vulnerabilities. While ABL falters with $75.45\%$ ASR against BadNets and $90.39\%$ against Adap-Blend on CIFAR-10, CSC maintains an ASR under $0.1\%$ across all scenarios. It also sustains ASRs below $0.5\%$ against sophisticated feature-space attacks like WaNet and IAB on GTSRB. As indicated in \autoref{tab: main results resnet18}, these outcomes validate the robust, scalable, and generalized efficacy of CSC in thwarting diverse poisoning-based threats.

\begin{table}[!t]
\centering
\caption{Comparison results of different poisoned sample segregation methods.}
\resizebox{\linewidth}{!}{
\begin{tabular}{l|cc|cc|cc|cc|cc|cc}
\toprule
{Defense $\rightarrow$} &\multicolumn{2}{c|}{SS} & \multicolumn{2}{c|}{AC} & \multicolumn{2}{c|}{ABL} & \multicolumn{2}{c|}{NONE} & \multicolumn{2}{c|}{D-ST} & \multicolumn{2}{c}{Ours} \\
\cline{1-13}
{Attack $\downarrow$} & Precision & Recall & Precision & Recall & Precision & Recall & Precision & Recall & Precision & Recall & Precision & Recall \\
\hline
Badnets & 13.7 & 38.2 & 37.2 & 80.9 & 1.9 & 0.2 & 25.7 & 85.4 & 1.3 & 0.8 & 100 & 94.9 \\
Trojan & 18.7 & 51.2 & 11.3 & 45.9 & 100 & 12.6 & 99.9 & 96.7 & 100 & 57.5 & 100 & 97.2 \\
Blend-Strip & 20.7 & 65.2 & 25.5 & 97.6 & 94.6 & 12.6 & 6.5 & 17.4 & 97.5 & 55.8 & 100 & 99.8 \\
Blend-Kitty & 20.9 & 65.2 & 23.7 & 100 & 94.1 & 12.1 & 17.6 & 98.4 & 100 & 57.6 & 98.9 & 100 \\
SIG & 1.0 & 31.2 & 5.4 & 57.4 & 37.4 & 36.8 & 0.1 & 4.7 & 3.9 & 18.7 & 100 & 94.2 \\
CL & 1.8 & 9.7 & 1.2 & 14.3 & 68.8 & 69.1 & 8.3 & 100 & 5.9 & 20.1 & 100 & 99.3 \\
WaNet & 4.3 & 12.6 & 24.5 & 96.5 & 30.2 & 36.1 & 51.8 & 9.4 & 6.8 & 4.2 & 98.5 & 96.4 \\
ABS & 18.5 & 57.3 & 20.1 & 98.5 & 99.2 & 9.4 & 27.3 & 8.6 & 7.8 & 6.3 & 100 & 95.8 \\
IAB & 14.8 & 45.1 & 24.7 & 99.4 & 100 & 11.7 & 97.6 & 8.9 & 8.8 & 4.7 & 100 & 98.2 \\
Adap-Blend & 17.3 & 64.5 & 61.3 & 81.2 & 3.5 & 2.7 & 6.6 & 2.4 & 4.9 & 7.6 & 98.2 & 84.6 \\
\bottomrule
\end{tabular}
}
\label{tab: poisoned sample sergregatin}
\end{table}

\subsection{Analysis of Poisoned Sample Segregation}
We initiate our assessment with a benchmarking of the poisoned sample segregation module against established methods.
Consistent with previous research, we incorporate two detection-oriented defenses as reference points: Spectral Signatures (SS) \cite{10.5555/3327757.3327896} and Activation Clustering (AC) \cite{DBLP:journals/corr/abs-1811-03728}.
Evaluations are performed on the CIFAR-10 across 10 distinct attacks, with  \autoref{tab: poisoned sample sergregatin} detailing the segregation efficacy of the examined methods.

\autoref{tab: poisoned sample sergregatin} demonstrates that CSC surpasses the baseline approaches SS and AC, attaining precision exceeding $98\%$ and recall above $94\%$ across all cases except Adap-Blend. Notably, CSC delivers substantial precision gains relative to AC, ranging from $36.9\%$ (improving from $61.3\%$ to $98.2\%$ for Adap-Blend) to $98.8\%$ (improving from $1.2\%$ to $100\%$ for CL).
Although CSC yields slightly lower recall than AC in the ABS and IAB scenarios (by $2.7\%$ and $1.2\%$, respectively), it achieves marked precision improvements, increasing rates from $20.1\%$ to $100\%$ and from $24.1\%$ to $100\%$.
This advantage stems from the capacity of CSC to recognize the tendency of benign samples to aggregate into dominant clusters early in training, thereby minimizing the misclassification of benign data. While SS and AC perform adequately against dirty-label and feature-space attacks, they fail against clean-label variants. In contrast, CSC remains highly effective across both dirty-label and clean-label threats.
For the SIG attack, CSC achieves $100\%$ precision and increases recall to $94.2\%$, compared to $31.2\%$ and $57.4\%$ for SS and AC, respectively.Unlike other defenses, CSC achieves high precision and recall simultaneously. Although ABL and D-ST reach $100\%$ precision, they exhibit constrained recall rates that peak at $69.1\%$ and $57.6\%$, respectively.
In comparison, CSC sustains nearly $100\%$ precision alongside recall exceeding $94\%$ across almost every attack scenario. NONE demonstrates strong performance against most dirty-label attacks but struggles to separate poisoned samples in clean-label and feature-space contexts.
For example, NONE identifies every poisoned sample in the CL attack to yield $100\%$ recall; however, its precision drops to $8.3\%$, indicating the erroneous exclusion of numerous benign samples.
These results highlight the limited adaptability of existing approaches. Conversely, CSC effectively accommodates diverse attacks and segregates poisoned samples with high reliability.

\subsection{Analysis of Backdoor Concealment}

\begin{table}[!t]
\scriptsize
\centering
\caption{ACC and ASR under four different backdoor removal methods. U.R. indicates Unlearning and Relearning}
\label{tab:acc_asr_under_four_backdoor_removal_methods}
\begin{tabular}{l|cc|cc|cc|cc|cc}
\toprule
Methods & \multicolumn{2}{c|}{BadNets} & \multicolumn{2}{c|}{Blend-Strip} & \multicolumn{2}{c|}{SIG} & \multicolumn{2}{c|}{CL} & \multicolumn{2}{c}{\textit{Avg.}} \\
& ACC & ASR & ACC & ASR & ACC & ASR & ACC & ASR & ACC & ASR \\
\hline
Unlearning & 82.45 & 0.0 & 83.17 & 0.0 & 49.62 & 0.02 & 83.71 & 0.0 & 74.74 & 0.005 \\
U.R. & 91.17 & 0.37 & 93.52 & 0.21 & 55.61 & 1.05 & 84.79 & 0.67 & 81.27 & 0.58 \\
Mix CE & 92.54 & 0.0 & 92.03 & 0.01 & 93.31 & 0.27 & 91.41 & 3.27 & 92.30 & 0.89 \\
Ours (BC) & 95.41 & 0.0 & 94.10 & 0.01 & 95.74 & 0.05 & 94.85 & 0.0 & 95.03 & 0.015 \\
\bottomrule
\end{tabular}
\end{table}

To evaluate Backdoor Concealment (BC), we benchmarked our approach against three established backdoor removal techniques: Unlearning, Unlearning \& Relearning (U.R.), and Mixed Cross-Entropy (Mix CE), implemented in ABL, D-BR, and D-ST, respectively. NONE, CBD, and ASD were omitted due to their incompatibility with our framework. We measured defensive effectiveness by integrating diverse erasure strategies into our segregation method. \autoref{tab:acc_asr_under_four_backdoor_removal_methods} depicts the eradication results for CSC compared to the alternative approaches.Although unlearning decreases ASR to $0\%$ across all attacks, it significantly degrades ACC. Conversely, BC improves ACC from $12.96\%$ (increasing from $82.45\%$ to $95.41\%$ for BadNets) to $46.12\%$ (increasing from $49.62\%$ to $95.74\%$ for SIG) over Unlearning, while achieving equivalent backdoor eradication. These results highlight the capacity of BC to eliminate backdoors with negligible impact on model utility.Furthermore, Unlearning risks model collapse. As shown in \autoref{tab:acc_asr_under_four_backdoor_removal_methods}, Unlearning and U.R. yield ACCs of only $49.62\%$ and $55.61\%$, respectively, against the SIG attack. In contrast, BC attains $95.74\%$ ACC in the same scenario. While Mix CE balances ACC maintenance and ASR reduction, our approach outperforms it by improving average ACC by $2.73\%$ and lowering average ASR from $0.89\%$ to $0.015\%$. Overall, these experiments confirm that our method preserves model accuracy while effectively suppressing ASR.

\begin{figure}[!t]
\centering
\includegraphics[width=0.65\linewidth]{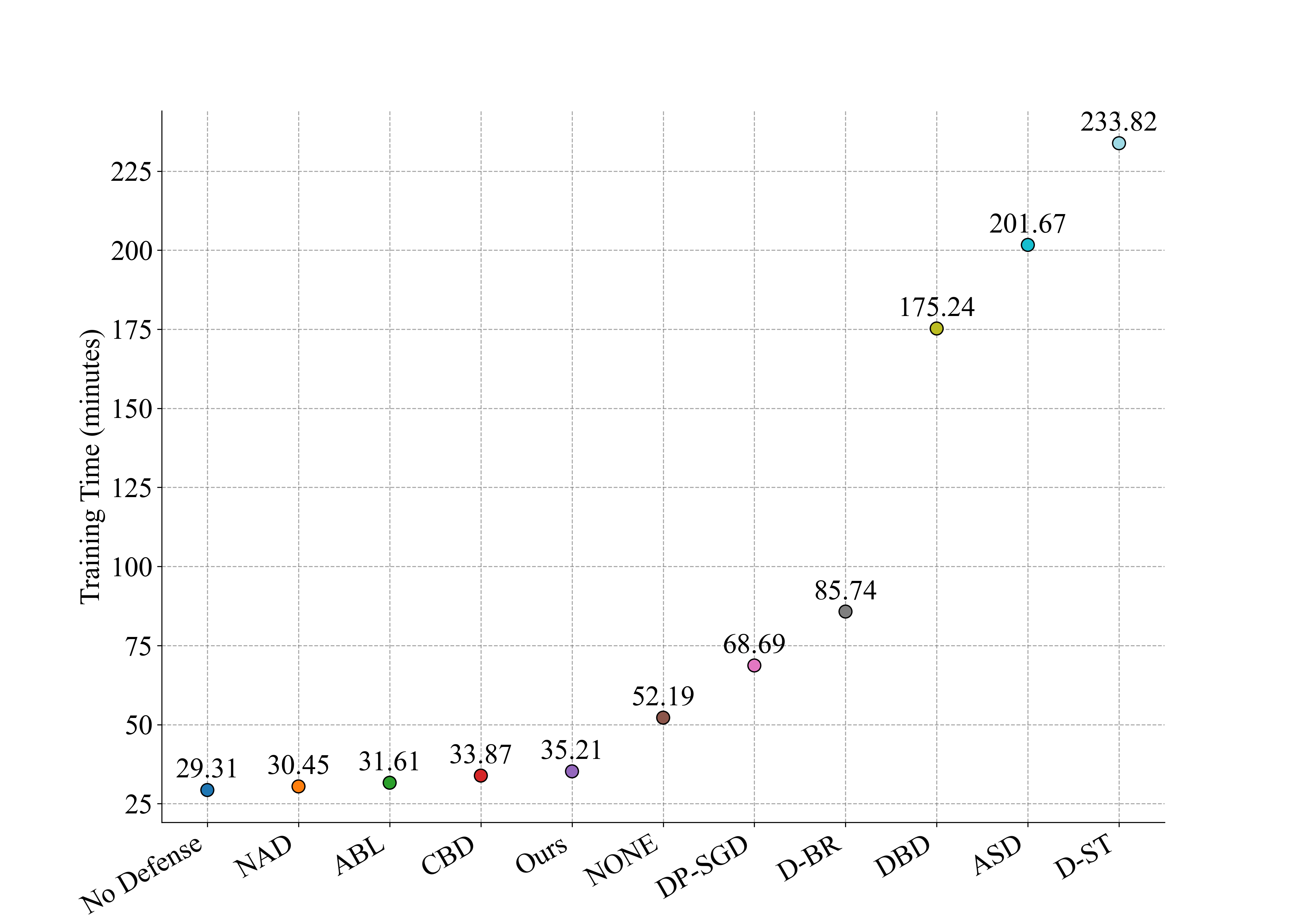}
\caption{The time consumption of BadNets as an attack method on the CIFAR-10 dataset across different defense methods.}
\label{fig: training time}
\end{figure}

\subsection{Analysis on Training Time}
We empirically evaluates the computational efficiency of CSC.
Specifically, we measure the time required by CSC, nine backdoor defenses, and an undefended baseline to counter BadNets on CIFAR-10. \autoref{fig: training time} shows that CSC requires slightly more time than undefended training (35.21 versus 29.31 minutes) due to additional components for segregating poisoned samples and concealing backdoors. However, this overhead is significantly lower than that of NONE, DP-SGD, D-BR, DBD, ASD, and D-ST, and it remains comparable to NAD, ABL, and CBD. Furthermore, the previous section demonstrates the superior effectiveness of CSC over prior methods. Because training typically occurs offline, this modest increase in computational time is acceptable.

\subsection{Different Poisoning Rate}
\begin{table}[t]
\centering
\caption{The ASR and ACC of CSC on CIFAR-10 under different poisoning rates $\gamma$.}
\label{tab:poisoning_rate}
\resizebox{\linewidth}{!}{
\begin{tabular}{l|cc|cc|cc|cc|cc|cc}
\toprule
Poisoning Rate & \multicolumn{4}{c|}{ $\gamma = 1\%$} & \multicolumn{4}{c|}{$\gamma=5\%$} & \multicolumn{4}{c}{$\gamma = 10\%$} \\
\cmidrule(lr){2-5}\cmidrule(lr){6-9}\cmidrule(lr){10-13}
Methods & \multicolumn{2}{c|}{No Defense} & \multicolumn{2}{c|}{Ours} & \multicolumn{2}{c|}{No Defense} & \multicolumn{2}{c|}{Ours} & \multicolumn{2}{c|}{No Defense} & \multicolumn{2}{c}{Ours} \\
 & ACC & ASR & ACC & ASR & ACC & ASR & ACC & ASR & ACC & ASR & ACC & ASR \\
\midrule
BadNets & 95.72 & 98.91 & 94.89 & 0.0 & 94.56 & 100 & 94.15 & 0.0 & 93.53 & 100 & 93.71 & 0.01 \\
Trojan & 95.38 & 98.22 & 95.12 & 0.03 & 94.71 & 99.99 & 94.73 & 0.02 & 94.51 & 100 & 94.18 & 0.01 \\
Blend-Strip & 95.12 & 97.64 & 94.47 & 0.01 & 94.38 & 100 & 94.49 & 0.01 & 94.26 & 100 & 94.27 & 0.00 \\
Blend-Kitty & 95.31 & 99.35 & 94.65 & 0.01 & 94.83 & 99.37  & 94.05 & 0.0 & 94.31 & 100 & 95.01 & 0.00 \\
\textit{Avg.} & 95.38 & 98.53 & 94.78 & 0.013 & 94.62 & 99.84 & 94.36 & 0.008 & 94.15 & 100 & 94.29 & 0.005 \\
\bottomrule
\end{tabular}
}
\end{table}

Different poisoning rates $\gamma$ may influence the performance of CSC.
In this paper, we evaluated $\gamma$ values of $1\%$, $5\%$ and $10\%$ to defend against a range of attacks.
The results for CSC across these rates are presented in \autoref{tab:poisoning_rate}, which indicates that CSC preserves elevated ACC and lowers ASR to near zero regardless of the poisoning rate.
Specifically, with the poisoning rate $\gamma$ set at $1\%$, CSC decreases the average ASR from $98.53\%$ to $0.013\%$ along with a modest average ACC reduction of $0.6\%$.
In a comparable manner, at $\gamma=5\%$, the method reduces ASR from $99.84\%$ to $0.008\%$ while incurring an average ACC drop of only $0.21\%$.
Overall, these outcomes confirm CSC's robust efficacy under diverse poisoning rates.

\begin{figure}[!t]
    \centering
    \begin{subfigure}[b]{0.48\textwidth}
        \centering
        \includegraphics[width=\textwidth]{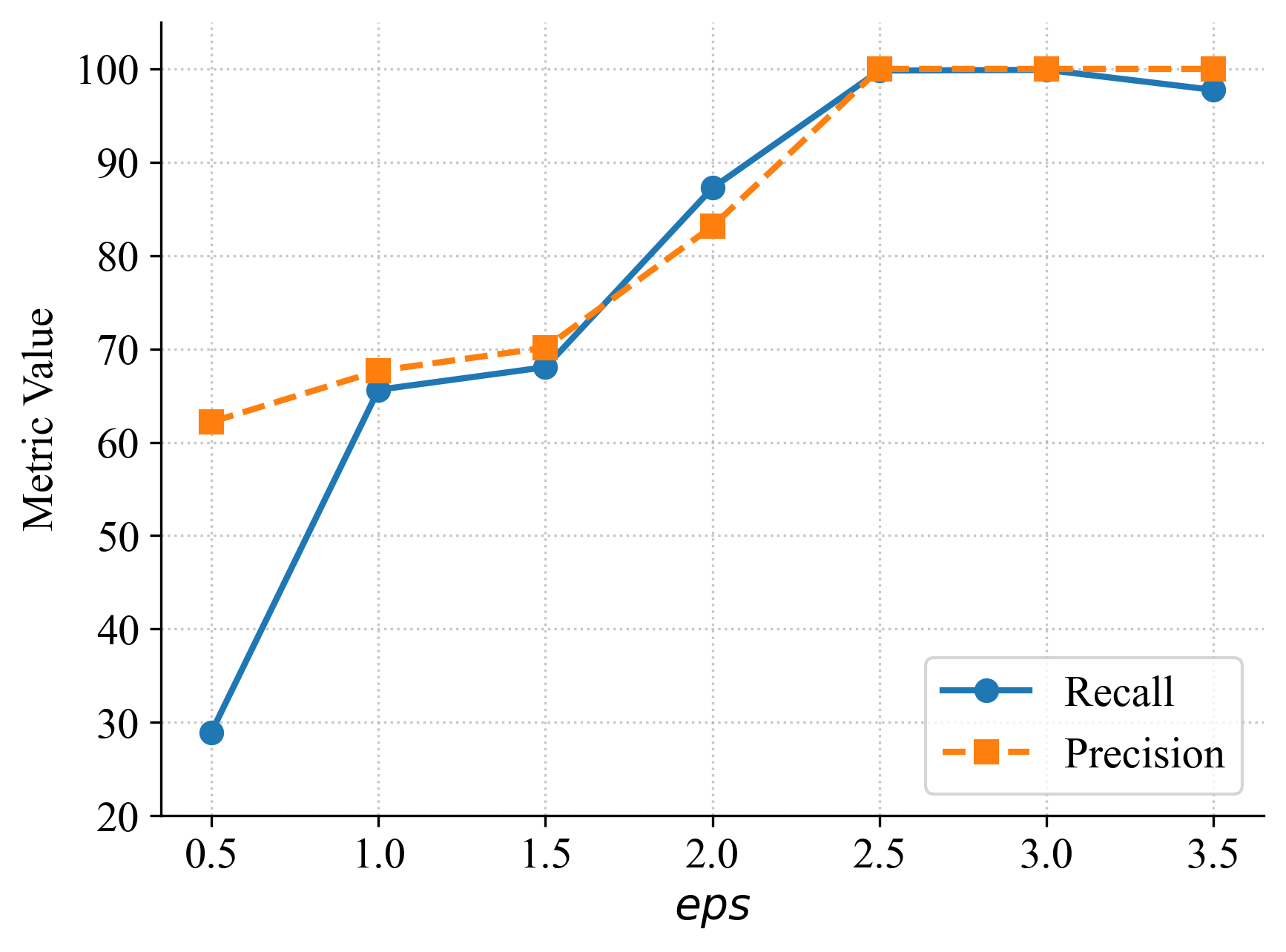}
        \caption{Impact of eps}
        \label{fig:dbscan_eps}
    \end{subfigure}
    \hfill 
    \begin{subfigure}[b]{0.48\textwidth}
        \centering
        \includegraphics[width=\textwidth]{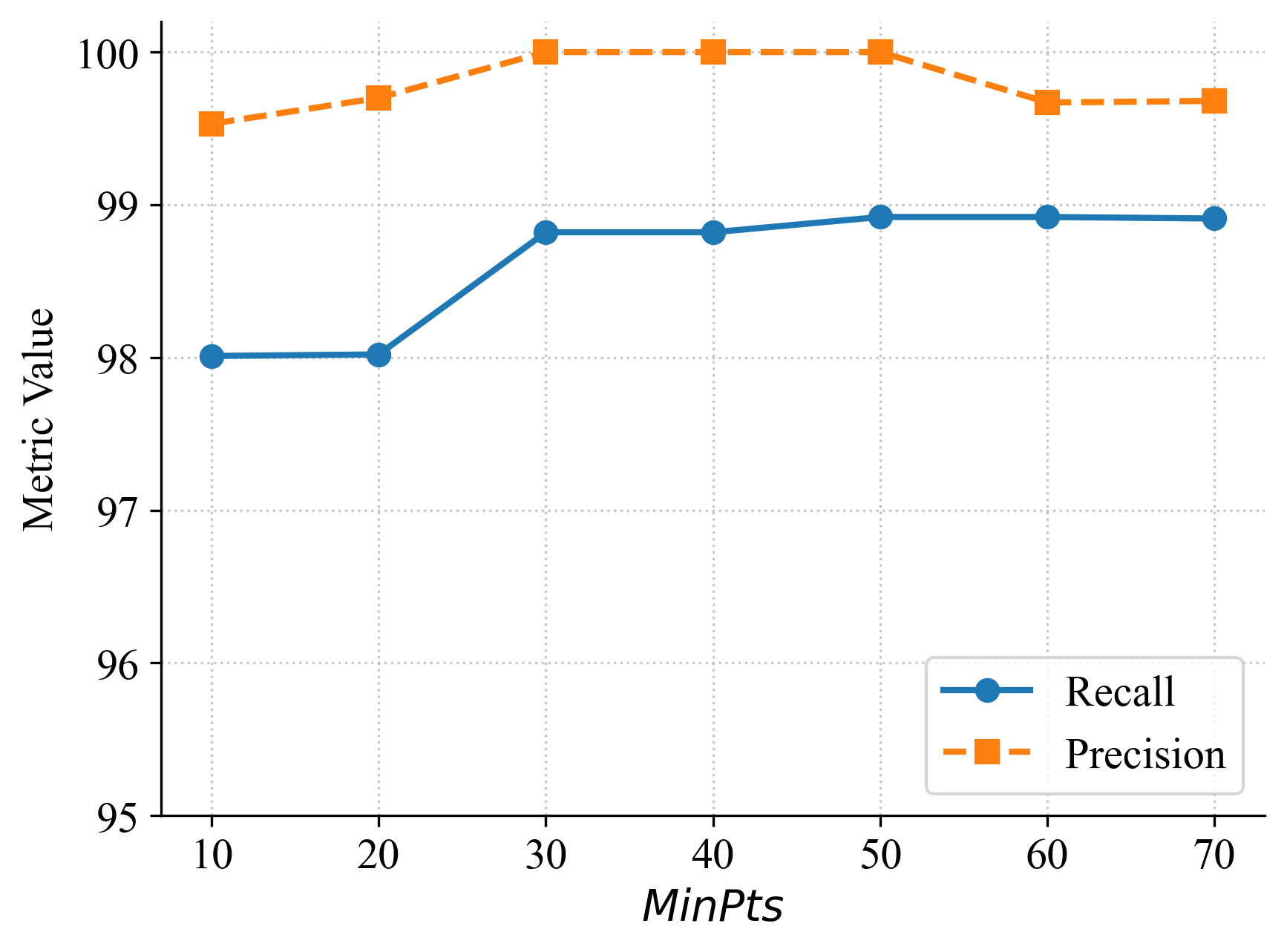}
        \caption{Impact of $MinPts$}
        \label{fig:dbscan_minpts}
    \end{subfigure}
    \caption{Evaluation of DBSCAN performance under different hyperparameters.}
    \label{fig:dbscan_hyperparameters}
\end{figure}

\subsection{Hyperparameters of DBSCAN}
To evaluate the impact of the DBSCAN hyperparameter settings, we conducted experiments on the CIFAR-10 dataset against attacks including Trojan, Blend, and SIG, utilizing various values for $eps$ and $MinPts$. 
The poisoning rate $\gamma$ was maintained at 10\%. 
When evaluating $eps$, we fixed $MinPts$ to its default value, and vice versa. \autoref{fig:dbscan_hyperparameters} illustrates the recall and precision of our poisoned sample segregation method across different $eps$ and $MinPts$ values.
As observed, an optimal balance between recall and precision is achieved when $eps$ is set to 3. Consequently, we adopted 3 as the default setting for $eps$ in our experiments. 
Furthermore, we found that our poisoned sample segregation approach demonstrates robustness to the $MinPts$ parameter, as the results exhibited minimal variance when the value ranged from 10 to 70.

\section{Conclusion}

Our study reexamines the fundamental principles underlying poisoning-based backdoor attacks, offering a detailed analysis of their dynamics within latent representations throughout model training.
Two primary findings emerge from our analysis: 1) during initial training stages, poisoned samples aggregate into separate clusters that are clearly segregated from benign samples; 2) upon the completion of training, the model's classification head establishes a distinct connection linking backdoor features directly to the intended target class.
Leveraging these insights, we introduce CSC, an innovative defense strategy that isolates poisoned samples from benign ones and eradicates the embedded backdoor.
Empirical assessments demonstrate that CSC substantially outperforms contemporary poison restraint-based defenses, achieving superior results consistently across diverse attacks and datasets.


%
%
%
\bibliographystyle{splncs04}
\bibliography{ref}

@article{kuutti2020survey,
  title={A survey of deep learning applications to autonomous vehicle control},
  author={Kuutti, Sampo and Bowden, Richard and Jin, Yaochu and Barber, Phil and Fallah, Saber},
  journal={IEEE Transactions on Intelligent Transportation Systems},
  volume={22},
  number={2},
  pages={712--733},
  year={2020},
  publisher={IEEE}
}

@article{shen2017deep,
  title={Deep learning in medical image analysis},
  author={Shen, Dinggang and Wu, Guorong and Suk, Heung-Il},
  journal={Annual review of biomedical engineering},
  volume={19},
  number={1},
  pages={221--248},
  year={2017},
  publisher={Annual Reviews}
}

@inproceedings{huang2022backdoor,
  title={Backdoor Defense via Decoupling the Training Process},
  author={Huang, Kunzhe and Li,  Yiming and Wu, Baoyuan  and Qin, Zhan and Ren, Kui},
  booktitle={International Conference on Learning Representations (ICLR)},
  year={2022}
}

@article{wang2022training,
  title={Training with more confidence: Mitigating injected and natural backdoors during training},
  author={Wang, Zhenting and Ding, Hailun and Zhai, Juan and Ma, Shiqing},
  journal={Advances in Neural Information Processing Systems},
  volume={35},
  pages={36396--36410},
  year={2022}
}

@inproceedings{wang2019neural,
  title={Neural cleanse: Identifying and mitigating backdoor attacks in neural networks},
  author={Wang, Bolun and Yao, Yuanshun and Shan, Shawn and Li, Huiying and Viswanath, Bimal and Zheng, Haitao and Zhao, Ben Y},
  booktitle={2019 IEEE symposium on security and privacy (SP)},
  pages={707--723},
  year={2019},
  organization={IEEE}
}

@inproceedings{zhu2019transferable,
  title={Transferable clean-label poisoning attacks on deep neural nets},
  author={Zhu, Chen and Huang, W Ronny and Li, Hengduo and Taylor, Gavin and Studer, Christoph and Goldstein, Tom},
  booktitle={International conference on machine learning},
  pages={7614--7623},
  year={2019},
  organization={PMLR}
}

@inproceedings{bourtoule2021machine,
  title={Machine unlearning},
  author={Bourtoule, Lucas and Chandrasekaran, Varun and Choquette-Choo, Christopher A and Jia, Hengrui and Travers, Adelin and Zhang, Baiwu and Lie, David and Papernot, Nicolas},
  booktitle={2021 IEEE symposium on security and privacy (SP)},
  pages={141--159},
  year={2021},
  organization={IEEE}
}

@article{10.1145/3603620,
author = {Xu, Heng and Zhu, Tianqing and Zhang, Lefeng and Zhou, Wanlei and Yu, Philip S.},
title = {Machine Unlearning: A Survey},
year = {2023},
issue_date = {January 2024},
publisher = {Association for Computing Machinery},
address = {New York, NY, USA},
volume = {56},
number = {1},
issn = {0360-0300},
url = {https://doi.org/10.1145/3603620},
doi = {10.1145/3603620},
journal = {ACM Comput. Surv.},
month = aug,
articleno = {9},
numpages = {36}
}

@article{gu2017badnets,
  title={Badnets: Identifying vulnerabilities in the machine learning model supply chain},
  author={Gu, Tianyu and Dolan-Gavitt, Brendan and Garg, Siddharth},
  journal={arXiv preprint arXiv:1708.06733},
  year={2017}
}

@inproceedings{liu2018trojaning,
  author    = {Yingqi Liu and
               Shiqing Ma and
               Yousra Aafer and
               Wen-Chuan Lee and
               Juan Zhai and
               Weihang Wang and
               Xiangyu Zhang},
  title     = {Trojaning Attack on Neural Networks},
  booktitle = {25th Annual Network and Distributed System Security Symposium, {NDSS}
               2018, San Diego, California, USA, February 18-221, 2018},
  publisher = {The Internet Society},
  year      = {2018},
}

@article{chen2017targeted,
  title={Targeted backdoor attacks on deep learning systems using data poisoning},
  author={Chen, Xinyun and Liu, Chang and Li, Bo and Lu, Kimberly and Song, Dawn},
  journal={arXiv preprint arXiv:1712.05526},
  year={2017}
}

@inproceedings{qi2023revisiting,
title={Revisiting the Assumption of Latent Separability for Backdoor Defenses},
author={Xiangyu Qi and Tinghao Xie and Yiming Li and Saeed Mahloujifar and Prateek Mittal},
booktitle={The Eleventh International Conference on Learning Representations },
year={2023},
url={https://openreview.net/forum?id=_wSHsgrVali}
}

@inproceedings{barni2019new,
  title={A new backdoor attack in cnns by training set corruption without label poisoning},
  author={Barni, Mauro and Kallas, Kassem and Tondi, Benedetta},
  booktitle={2019 IEEE International Conference on Image Processing (ICIP)},
  pages={101--105},
  year={2019},
  organization={IEEE}
}

@article{turner2019label,
  title={Label-consistent backdoor attacks},
  author={Turner, Alexander and Tsipras, Dimitris and Madry, Aleksander},
  journal={arXiv preprint arXiv:1912.02771},
  year={2019}
}

@article{nguyen2020input,
  title={Input-aware dynamic backdoor attack},
  author={Nguyen, Tuan Anh and Tran, Anh},
  journal={Advances in Neural Information Processing Systems},
  volume={33},
  pages={3454--3464},
  year={2020}
}

@inproceedings{
nguyen2021wanet,
title={WaNet - Imperceptible Warping-based Backdoor Attack},
author={Tuan Anh Nguyen and Anh Tuan Tran},
booktitle={International Conference on Learning Representations},
year={2021},
url={https://openreview.net/forum?id=eEn8KTtJOx}
}

@inproceedings{
li2022baat,
title={{BAAT}: Towards Sample-specific Backdoor Attack with Clean Labels},
author={Yiming Li and Mingyan Zhu and Chengxiao Luo and Haiqin Weng and Yong Jiang and Tao Wei and Shu-Tao Xia},
booktitle={NeurIPS ML Safety Workshop},
year={2022},
url={https://openreview.net/forum?id=kwlkmbebcqP}
}

@INPROCEEDINGS{9711191,
  author={Li, Yuezun and Li, Yiming and Wu, Baoyuan and Li, Longkang and He, Ran and Lyu, Siwei},
  booktitle={2021 IEEE/CVF International Conference on Computer Vision (ICCV)}, 
  title={Invisible Backdoor Attack with Sample-Specific Triggers}, 
  year={2021},
  volume={},
  number={},
  pages={16443-16452},
  doi={10.1109/ICCV48922.2021.01615}}

@article{10.1145/3694965,
author = {Kombrink, Meike Helena and Geradts, Zeno Jean Marius Hubert and Worring, Marcel},
title = {Image Steganography Approaches and Their Detection Strategies: A Survey},
year = {2024},
issue_date = {February 2025},
publisher = {Association for Computing Machinery},
address = {New York, NY, USA},
volume = {57},
number = {2},
issn = {0360-0300},
url = {https://doi.org/10.1145/3694965},
doi = {10.1145/3694965},
journal = {ACM Comput. Surv.},
month = oct,
articleno = {33},
numpages = {40}
}

@inproceedings{10.1145/3319535.3363216,
author = {Liu, Yingqi and Lee, Wen-Chuan and Tao, Guanhong and Ma, Shiqing and Aafer, Yousra and Zhang, Xiangyu},
title = {ABS: Scanning Neural Networks for Back-doors by Artificial Brain Stimulation},
year = {2019},
isbn = {9781450367479},
publisher = {Association for Computing Machinery},
address = {New York, NY, USA},
url = {https://doi.org/10.1145/3319535.3363216},
doi = {10.1145/3319535.3363216},
booktitle = {Proceedings of the 2019 ACM SIGSAC Conference on Computer and Communications Security},
pages = {1265–1282},
numpages = {18},
location = {London, United Kingdom},
series = {CCS '19}
}

@inproceedings{10.5555/3001460.3001507,
author = {Ester, Martin and Kriegel, Hans-Peter and Sander, J\"{o}rg and Xu, Xiaowei},
title = {A density-based algorithm for discovering clusters in large spatial databases with noise},
year = {1996},
publisher = {AAAI Press},
booktitle = {Proceedings of the Second International Conference on Knowledge Discovery and Data Mining},
pages = {226–231},
numpages = {6},
location = {Portland, Oregon},
series = {KDD'96}
}

@inproceedings{Krizhevsky2009LearningML,
  title={Learning Multiple Layers of Features from Tiny Images},
  author={Alex Krizhevsky},
  year={2009},
  url={https://api.semanticscholar.org/CorpusID:18268744}
}

@inproceedings{he2016deep,
  title={Deep residual learning for image recognition},
  author={He, Kaiming and Zhang, Xiangyu and Ren, Shaoqing and Sun, Jian},
  booktitle={Proceedings of the IEEE Conference on Computer Vision and Pattern Recognition (CVPR)},
  pages={770--778},
  year={2016}
}

@article{Stallkamp2012,
title = "Man vs. computer: Benchmarking machine learning algorithms for traffic sign recognition",
journal = "Neural Networks",
volume = "",
number = "0",
pages = " - ",
year = "2012",
note = "",
issn = "0893-6080",
doi = "10.1016/j.neunet.2012.02.016",
url = "http://www.sciencedirect.com/science/article/pii/S0893608012000457",
author = "J. Stallkamp and M. Schlipsing and J. Salmen and C. Igel"
}

@inproceedings{deng2009imagenet,
  title={Imagenet: A large-scale hierarchical image database},
  author={Deng, Jia and Dong, Wei and Socher, Richard and Li, Li-Jia and Li, Kai and Fei-Fei, Li},
  booktitle={2009 IEEE conference on computer vision and pattern recognition},
  pages={248--255},
  year={2009},
  organization={IEEE}
}

@article{Hong2020OnTE,
  title={On the Effectiveness of Mitigating Data Poisoning Attacks with Gradient Shaping},
  author={Sanghyun Hong and Varun Chandrasekaran and Yigitcan Kaya and Tudor Dumitras and Nicolas Papernot},
  journal={ArXiv},
  year={2020},
  volume={abs/2002.11497},
  url={https://api.semanticscholar.org/CorpusID:211506328}
}

@inproceedings{
li2021neural,
title={Neural Attention Distillation: Erasing Backdoor Triggers from Deep Neural Networks},
author={Yige Li and Xixiang Lyu and Nodens Koren and Lingjuan Lyu and Bo Li and Xingjun Ma},
booktitle={International Conference on Learning Representations},
year={2021},
url={https://openreview.net/forum?id=9l0K4OM-oXE}
}

@inproceedings{
li2021antibackdoor,
title={Anti-Backdoor Learning: Training Clean Models on Poisoned Data},
author={Yige Li and Xixiang Lyu and Nodens Koren and Lingjuan Lyu and Bo Li and Xingjun Ma},
booktitle={Advances in Neural Information Processing Systems},
editor={A. Beygelzimer and Y. Dauphin and P. Liang and J. Wortman Vaughan},
year={2021},
url={https://openreview.net/forum?id=cAw860ncLRW}
}

@inproceedings{
chen2022effective,
title={Effective Backdoor Defense by Exploiting Sensitivity of Poisoned Samples},
author={Weixin Chen and Baoyuan Wu and Haoqian Wang},
booktitle={Advances in Neural Information Processing Systems},
editor={Alice H. Oh and Alekh Agarwal and Danielle Belgrave and Kyunghyun Cho},
year={2022},
url={https://openreview.net/forum?id=AsH-Tx2U0Ug}
}

@INPROCEEDINGS{10204451,
  author={Zhang, Zaixi and Liu, Qi and Wang, Zhicai and Lu, Zepu and Hu, Qingyong},
  booktitle={2023 IEEE/CVF Conference on Computer Vision and Pattern Recognition (CVPR)}, 
  title={Backdoor Defense via Deconfounded Representation Learning}, 
  year={2023},
  volume={},
  number={},
  pages={12228-12238},
  doi={10.1109/CVPR52729.2023.01177}}

@INPROCEEDINGS{10204454,
  author={Gao, Kuofeng and Bai, Yang and Gu, Jindong and Yang, Yong and Xia, Shu-Tao},
  booktitle={2023 IEEE/CVF Conference on Computer Vision and Pattern Recognition (CVPR)}, 
  title={Backdoor Defense via Adaptively Splitting Poisoned Dataset}, 
  year={2023},
  volume={},
  number={},
  pages={4005-4014},
  doi={10.1109/CVPR52729.2023.00390}}

@inproceedings{10.5555/3327757.3327896,
author = {Tran, Brandon and Li, Jerry and M\k{a}dry, Aleksander},
title = {Spectral signatures in backdoor attacks},
year = {2018},
publisher = {Curran Associates Inc.},
address = {Red Hook, NY, USA},
booktitle = {Proceedings of the 32nd International Conference on Neural Information Processing Systems},
pages = {8011–8021},
numpages = {11},
location = {Montr\'{e}al, Canada},
series = {NIPS'18}
}

@article{DBLP:journals/corr/abs-1811-03728,
  author       = {Bryant Chen and
                  Wilka Carvalho and
                  Nathalie Baracaldo and
                  Heiko Ludwig and
                  Benjamin Edwards and
                  Taesung Lee and
                  Ian M. Molloy and
                  Biplav Srivastava},
  title        = {Detecting Backdoor Attacks on Deep Neural Networks by Activation Clustering},
  journal      = {CoRR},
  volume       = {abs/1811.03728},
  year         = {2018},
  url          = {http://arxiv.org/abs/1811.03728},
  eprinttype    = {arXiv},
  eprint       = {1811.03728},
  bibsource    = {dblp computer science bibliography, https://dblp.org}
}

@article{chen2026turning,
  title={Turning Black Box into White Box: Dataset Distillation Leaks},
  author={Chen, Huajie and Zhu, Tianqing and Zhong, Yuchen and Zhang, Yang and Wang, Shang and He, Feng and Zhang, Lefeng and Shen, Jialiang and Wang, Minghao and Zhou, Wanlei},
  journal={arXiv preprint arXiv:2603.01053},
  year={2026}
}

@article{chen2025queen,
  title={Queen: Query unlearning against model extraction},
  author={Chen, Huajie and Zhu, Tianqing and Zhang, Lefeng and Liu, Bo and Wang, Derui and Zhou, Wanlei and Xue, Minhui},
  journal={IEEE Transactions on Information Forensics and Security},
  year={2025},
  publisher={IEEE}
}

@article{chen2024high,
  title={High-frequency matters: attack and defense for image-processing model watermarking},
  author={Chen, Huajie and Zhu, Tianqing and Liu, Chi and Yu, Shui and Zhou, Wanlei},
  journal={IEEE Transactions on Services Computing},
  volume={17},
  number={4},
  pages={1565--1579},
  year={2024},
  publisher={IEEE}
}

@article{shi2026osmosis,
  title={Osmosis Distillation: Model Hijacking with the Fewest Samples},
  author={Shi, Yuchen and Chen, Huajie and Xu, Heng and Liu, Zhiquan and Shen, Jialiang and Liu, Chi and Zhou, Shuai and Zhu, Tianqing and Zhou, Wanlei},
  journal={arXiv preprint arXiv:2603.04859},
  year={2026}
}

@article{chen2026hide,
  title={Hide\&Seek: Remove Image Watermarks with Negligible Cost via Pixel-wise Reconstruction},
  author={Chen, Huajie and Zhu, Tianqing and Yang, Hailin and Zhong, Yuchen and Zhang, Yang and Sun, Hui and Xu, Heng and Ying, Zuobin and Yin, Lihua and Zhou, Wanlei},
  journal={arXiv preprint arXiv:2603.01067},
  year={2026}
}

\end{document}